\documentclass[a4paper,12pt]{article}
\pdfoutput=1
\usepackage{graphicx}
\usepackage{amsthm}
\usepackage{amsfonts}
\usepackage{amsmath,latexsym}
\usepackage{amssymb}
\usepackage{hyperref}
\usepackage{tabularx}
\newcolumntype{M}{>{$}c<{$}}
\usepackage[matrix,arrow,curve]{xy}
\usepackage{placeins}
\usepackage{dsfont}

\numberwithin{equation}{section} \numberwithin{figure}{section}
\numberwithin{table}{section}

\def\papertitlepage{\baselineskip 3.5ex\thispagestyle{empty}}
\def\Title#1{\baselineskip 1cm \vspace{1.5cm}%
  \begin{center}{\Large\bf #1}\end{center}\vspace{0.5cm}}
\def\Authors#1{\begin{center}\renewcommand{\thefootnote}{\fnsymbol{footnote}}{\it #1}\end{center}}
\def\Abstract{\vspace{1.0cm}%
  \begin{center}{\large\bf Abstract}\end{center}}

\renewenvironment{thebibliography}{\pagebreak[3]\par\vspace{0.6em}
\begin{flushleft}{\large \bf References}\end{flushleft}
\vspace{-1.0em}

\begin{enumerate}\if@twocolumn\baselineskip=0.6em\itemsep -0.2em
\else\itemsep -0.2em\fi\labelsep 0.1em}{\end{enumerate} }

\def \rar {\rightarrow}
\def \la {\langle}
\def \ra {\rangle}
\def \del {\partial}
\def \Id {\mathds{1}}
\def \inf {\infty}
\def \ps {\phantom{-}}
\def \nn {\nonumber}

\begin{document}
{\papertitlepage \vspace*{0cm} {\hfill
\begin{minipage}{4.2cm}
%IFT-P. 2010\par\noindent
CCNH-UFABC 2019\par\noindent  November, 2019
\end{minipage}}
%%%%%%%%%%%%%%%%%%%%%%%%%%%%
%\vspace{1cm}
\Title{Numerical solution for tachyon vacuum in the Schnabl gauge}
%\vspace{1cm}
\Authors{{\sc E.~Aldo~Arroyo${}^{1}$\footnote{\tt
aldo.arroyo@ufabc.edu.br}}, and {\sc Mat\v{e}j Kudrna
${}^2$\footnote{{\tt kudrnam@fzu.cz}}}
\\
${}^1$Centro de Ci\^{e}ncias Naturais e Humanas, Universidade Federal do ABC \\[-2ex]
Santo Andr\'{e}, 09210-170 S\~{a}o Paulo, SP, Brazil
\\
${}^2$Institute of Physics of the ASCR, v.v.i. \\[-2ex]
Na Slovance 2, 182 21 Prague 8, Czech Republic }
} % \papertitlepage

\vskip-\baselineskip
%%%%%%%%%%%%%%%%%%%%%%%%%%%%%%%%%%%%%%%%%%%%%%%%%%%%%%%%%%%%%%%%
{\baselineskip .5cm \Abstract Based on the level truncation
scheme, we develop a new numerical method to evaluate the tachyon
vacuum solution in the Schnabl gauge up to level $L=24$. We
confirm the prediction that the energy associated to this
numerical solution has a local minimum at level $L=12$.
Extrapolating the energy data of $L \leq 24$ to infinite level, we
observe that the energy goes towards the analytical value $-1$,
nevertheless the precision of the extrapolation is lower than in
the Siegel gauge. Furthermore, we analyze the Ellwood invariant
and show that its value converges monotonically towards the
expected analytical result. We also study the tachyon vacuum
expectation value (vev) and some other coefficients of the
solution. Finally, some consistency checks of the solution are
performed, and we briefly discuss the search for other Schnabl
gauge numerical solutions.
%\\
%\\
%\textbf{Keywords}: String field theory, tachyon condensation,
%Pad\'{e} approximation.
 }
\newpage
\setcounter{footnote}{0}
%%%%%%%%%%%%%%%%%%%%%%%%%%%%%%%%%%%%%%%%%%%%%%%%%%%%%%%%%%%%%%%%
\tableofcontents

\section{Introduction}\label{sec:intro}
In the context of Witten's open bosonic string field theory
\cite{Witten:1985cc}, using a solution for tachyon condensation in
the Schnabl gauge \cite{Schnabl:2005gv}, the first analytic proof
of Sen's first conjecture \cite{Sen:1999mh,Sen:1999xm} has been
performed. Based on Schnabl's work, many analytical and numerical
tools have been developed
\cite{Okawa:2006vm,Okawa:2012ica,Fuchs:2006an,Arroyo:2011zt,Rastelli:2006ap,Ellwood:2006ba,
Okawa:2006sn,Erler:2006hw,Erler:2006ww,Schnabl:2010tb,Fuchs:2008cc,Kiermaier:2007jg,
Takahashi:2007du,Kishimoto:2011zza,Arroyo:2009ec,AldoArroyo:2011gx,Arroyo:2010sy}
and the techniques have been employed in the construction and
analysis of new analytical solutions
\cite{Takahashi:2002ez,Erler:2009uj,Murata:2011ep,Masuda:2012cj,Hata:2012cy,Masuda:2012kt,
Bonora:2011ru,Bonora:2011ri,Kiermaier:2007ba,Mertes:2016vos,Jokel:2017vlt,Arroyo:2017mpd}.

In the construction of this analytical solution for tachyon
condensation, the Schnabl gauge condition, ${\cal B}_0 \Psi=0$,
plays a fundamental role. As shown in reference
\cite{Schnabl:2005gv}, thanks to the combination of the ${\cal
B}_0$ gauge with the ${\cal L}_0$ level truncation in certain
sectors of the state space formed by the $\tilde c_n$ modes,
${\cal L}_0 + {\cal L}_0^\dagger$ and ${\cal B}_0 + {\cal
B}_0^\dagger$ operators acting on the vacuum, the entire set of
equations of motion $Q\Psi+\Psi*\Psi=0$ can be solved exactly, in
a recursive way. The result of such a calculation gives us the
analytical solution $\Psi$, which, in terms of wedge states with
ghost insertions, can be written as
\begin{eqnarray}
\label{intro4} \Psi &=& \lim_{N \rightarrow \infty} \Big[ \psi_N-
\sum_{n=0}^{N}
\partial_n \psi_n \Big] \; , \\
\label{intro5} \psi_n &=& \frac{2}{\pi^2} U^\dag_{n+2}U_{n+2}
\big[ (\mathcal{B}_0+\mathcal{B}^\dag_0)\tilde
c(-\frac{\pi}{4}n)\tilde c(\frac{\pi}{4}n) +\frac{\pi}{2} (\tilde
c(-\frac{\pi}{4}n) + \tilde c(\frac{\pi}{4}n)) \big] | 0\rangle \,
,
\end{eqnarray}
where $\psi_N$, with $N\rightarrow\infty$, is called the phantom
term \cite{Schnabl:2005gv,Okawa:2006vm,Erler:2006hw,Erler:2006ww}.
The equations (\ref{intro4}) and (\ref{intro5}) can be used to
expand the analytical solution in the state space of Virasoro
$L_0$ eigenstates. This level expansion of the solution is very
useful for the numerical evaluation of the energy. It is important
to mention that one slight disadvantage of the ${\cal B}_0$ gauge
is that the gauge fixing condition is broken by the Virasoro $L_0$
level truncation. In reference \cite{Schnabl:2005gv}, the author
conjectured that the level dependent Schnabl gauge fixing
condition would not pose problems and using the high $L_0$ level
truncation computations of Moeller and Taylor
\cite{Moeller:2000xv} and Gaiotto and Rastelli
\cite{Gaiotto:2002wy}, it should be possible to construct a
numerical solution that would converge to his analytical solution
when the level goes to infinity.

The first attempt to obtain a numerical solution for tachyon
vacuum in the Schnabl gauge was made by Arroyo et al.
\cite{Arroyo:2017iis}, using the traditional level truncation
computations up to level $L=10$. By extrapolating the energy data
of levels $L \leq 10$, shown in table \ref{Tabpaperaldo}, to
estimate the energy for $L> 10$, the authors predicted that the
energy reaches a local minimum value at level $L = 12$, to
subsequently turn back to approach $-1$ asymptotically as $L
\rightarrow \infty$. Although the value of the energy for this
numerical solution appears to converge to the expected analytical
result, the issue whether this solution could be identified with
the Schnabl analytical solution \cite{Schnabl:2005gv} when $L
\rightarrow \infty$ was inconclusive. For instance, as shown in
table \ref{Tabpaperaldo}, the tachyon vev (starting at level
$L=4$) appears to decrease with the level and it does not appear
to converge to the expected analytical value $t=0.55346558$.
Extrapolating the data of the tachyon vev \ref{Tabpaperaldo}, it
was predicted that the tachyon vev reaches a local minimum value
at a level close to $L=26$, to then turn back to approach the
expected analytical result as $L \rightarrow \infty$.

\begin{table}[ht]
\centering
\begin{tabular}{|c|c|c|}\hline
$L$ & $c_1 |0\rangle$  & $E^{Sch}$ \\ \hline 0   & 0.456177990470
& $-$0.684616159915 \\ \hline 2   & 0.544204232320 &
$-$0.959376599521 \\ \hline 4   & 0.548938521247 &
$-$0.994651904750 \\ \hline 6   & 0.548315148955 &
$-$1.003983765388 \\ \hline 8   & 0.547321883647 &
$-$1.007110280219 \\ \hline 10  & 0.546508411314 &
$-$1.008189759705 \\ \hline
\end{tabular}
\caption{$(L,3L)$ level truncation results of reference
\cite{Arroyo:2017iis} for the tachyon vev and vacuum energy in the
Schnabl gauge up to level $L=10$.} \label{Tabpaperaldo}
\end{table}

One of the main motivations of this work is to provide a
conclusive evidence of the conjecture in reference
\cite{Schnabl:2005gv}, that the numerical solution constructed in
the Schnabl gauge by means of level truncation computations can be
identified with the analytical solution (\ref{intro4}) when $L
\rightarrow \infty$. An obvious step to accomplish this task is to
perform higher level computations, this might appear as an
straightforward extension of the calculations developed in
reference \cite{Arroyo:2017iis}.

However the numerical method used in reference
\cite{Arroyo:2017iis} is not practical for levels beyond $L>12$.
To see why, let us briefly explain how it works. After truncating
the string field to some given level $L$, we plug this string
field into the string field theory action and compute the level
$(L,3L)$ tachyon potential. Then we impose the Schnabl gauge
condition and, to obtain the numerical solution, we extremize this
gauge fixed potential. Therefore this method needs the full
$(L,3L)$ level truncated potential as an input, however, storing
this full potential at high levels requires a huge amount of
computer memory, for example, to reproduce the level 24 results in
this way we would need a memory size over one petabyte.

In this work, we have managed to solve the aforementioned
technical issues, and we have obtained results up to level $L=24$,
using a clever numerical method based on the traditional level
truncation scheme, which in principle can be applied to all
general linear $b$-gauges. We have explicitly proven that the
energy of the numerical solution has in fact a local minimum at
level $L=12$, so the conjecture made in \cite{Arroyo:2017iis} was
proven to be correct.

Regarding the prediction of the local minimum at level $L=26$ for
the tachyon vev made in \cite{Arroyo:2017iis}, by extrapolating
the corresponding data of levels $L \leq 24$, we predict that this
local minimum probably exists, but at much higher level, possibly
around level $L=43$. Although this prediction is not very
reliable, from the trend observed in the data up to level 24, we
can state with certainty that this minimum does not happen at
$L=26$ as claimed in \cite{Arroyo:2017iis}.

Apart from the tachyon vev, we also analyzed the asymptotic
behavior of some other coefficients of the numerical solution, and
showed that they converge to the expected analytical result,
although the precision is lesser than we hoped for. Furthermore,
we computed the Ellwood invariant and found that its value is in
agreement with the expected analytical result. By performing some
consistency checks of the numerical solution, we provided an extra
evidence for the conclusion that the solution can be identified
with the analytical solution at the limit of infinite level.

This paper is organized as follows. In section 2, we discuss how
to impose the Schnabl gauge condition (or, in general, other
nontrivial linear gauge conditions) at high level numerical
calculations. We implement the gauge condition using a projector
acting on the full equations of motion. In section 3, we analyze
the tachyon vev and some other coefficients of the numerical
solution and compare them to coefficients of the analytical
solution. In section 4, we present level 24 data for the tachyon
vacuum energy and the gauge invariant overlap and we extrapolate
these quantities to the infinite level. In section 5, we check
that the numerical solution satisfies some nontrivial identities
that were discovered in \cite{Schnabl:2005gv}. In Section 6, we
verify that the solution satisfies some of the out-of-gauge
equations of motion. In section 7, we summarize our results and
discuss some related numerical experiments. In appendix A, we
provide general rules for extrapolations of various quantities to
infinite level. Finally, in appendix B, we briefly mention two
other numerical solutions in the Schnabl gauge.

\section{Solving the equations of motion in the Schnabl gauge}\label{sec:equationsgauge}
In this section, we discuss how to adapt Newton's method, which is
commonly used to solve the SFT equations numerically
\cite{Gaiotto:2002wy,KudrnaUniversal} with nontrivial gauge
conditions. More information about our numerical algorithms can be
found in \cite{KudrnaThesis}.

The string field theory action has a large amount of gauge
symmetry, which is, in an infinitesimal form, given by
\begin{equation}
\delta \Psi=Q\Lambda+(\Psi\ast\Lambda-\Lambda\ast\Psi).
\end{equation}
These gauge transformations do not commute with $L_0^{tot}$ and
therefore the gauge symmetry is broken when we truncate the action
to a finite level. This may look as an advantage because the
level-truncated equations of motion have only a discreet set of
solutions even without any gauge fixing, but it is actually the
opposite. The remnants of the gauge symmetry cause technical
problems and there does not appear to be any consistent way to
improve these solutions to higher levels (see
\cite{KudrnaUniversal}), which makes this approach essentially
unusable. Therefore, it is necessary to make a gauge choice in the
level truncation approach. We consider gauge conditions in the
form of a linear constraint,
\begin{equation}\label{gauge-fix}
\mathcal{G}\Psi=0.
\end{equation}
Ultimately, we are interested only in the Schnabl gauge, where
\begin{equation}
\mathcal{G}=\mathcal{B}_0=b_0+\sum_{k=1}^{\inf}\frac{2(-1)^{k+1}}{4k^2-1}
b_{2k},
\end{equation}
but the way to solve the equations of motion does not really
depend on the precise form of $\mathcal{G}$, so for now, we will
work with a generic linear operator $\mathcal{G}$.

Once we impose some gauge conditions, the system of the equations
of motion
\begin{equation}\label{equations full}
Q\Psi+\Psi\ast\Psi=0
\end{equation}
with the linear equations (\ref{gauge-fix}) becomes overdetermined
and it has generically no solutions at finite level. The usual
method to deal with this problem is to solve only a subset of the
full equations of motion, which we write as
\begin{equation}\label{equations solved}
P(Q\Psi+\Psi\ast \Psi)=0,
\end{equation}
where $P$ is a projector of an appropriate rank. The remaining
equations
\begin{equation}\label{equations unsolved}
(\Id-P)(Q\Psi+\Psi\ast \Psi)=0
\end{equation}
are left unsolved, but, for consistent solutions, they must
asymptotically approach zero as the level goes to infinity.
Following \cite{Gaiotto:2002wy}, we call them out-of-gauge
equations.

In the Siegel gauge, which is the most common choice in the level
truncation approach, these issues have a very elegant solution
\cite{Moeller:2000xv,Gaiotto:2002wy}. The gauge condition
\begin{equation}
b_0|\Psi\ra=0
\end{equation}
can be solved simply by removing all states that contain $c_0$
from the spectrum. The projected equations of motion are given by
derivatives of the action with respect to the remaining Siegel
gauge variables, which means that the projector is simply
$P=c_0b_0$.

However when we consider a more complicated gauge (which
essentially means any other gauge), such simple approach no longer
works and we will have to use the projector $P$ in a nontrivial
way.

We usually expand the string field as
\begin{equation}
|\Psi\ra=\sum_i t_i|i\ra,
\end{equation}
where $|i\ra$ are some basis states and $t_i$ is a vector of real
or complex coefficients. With respect to this basis, we define the
matrices
\begin{eqnarray}
Q_{ij} &=& \la i|Q|k\ra, \\
V_{ijk} &=& \la V_3|i\ra|j\ra|k\ra,
\end{eqnarray}
which allow us to write the level-truncated action as
\begin{eqnarray}\label{action truncated}
S(t)=-\frac{1}{g_o^2}\left(\frac{1}{2}\sum_{i,j}Q_{ij}t_it_j+\frac{1}{3}\sum_{i,j,k}V_{ijk}t_it_jt_k\right)
\end{eqnarray}
and the equations of motion as
\begin{equation}\label{EOM}
f_i(t)=\sum_{j}Q_{ij}t_j+\sum_{j,k}V_{ijk}t_jt_k=0.
\end{equation}
As long as there are no gauge conditions or they admit a trivial
solution like in the Siegel gauge, we solve these equations using
the well-known Newton's method. We start with an approximate
solution $t^{(0)}$ and we iteratively improve the solution as
$t^{(n+1)}=t^{(n)}+\Delta t$, where $\Delta t$ is a solution of
linear equations
\begin{equation}
\sum_j M_{ij}(t^{(n)})\Delta t_j=-f_i(t^{(n)}),
\end{equation}
where $M_{ij}$ is the Jacobian matrix
\begin{equation}
M_{ij}(t)=\frac{\del f_i(t)}{\del
t_j}=Q_{ij}+\sum_k(V_{ijk}+V_{jik})t_k.
\end{equation}
We repeat this procedure until we reach a desired precision of the
solution.

Now consider the gauge condition (\ref{gauge-fix}). Once we expand
the string fields into a basis, it transforms into a set of
homogeneous linear equations
\begin{equation}
\sum_j \mathcal{G}_{ij}t_j=0.
\end{equation}
Using standard linear algebra, we express the matrix $\mathcal{G}$
as\footnote{This form of the matrix is simplified for illustrative
purposes. We order states by level in the actual algorithm,
therefore the dependent and independent variables are not grouped
together and we end up with a matrix with permuted columns, see
the example (\ref{G example}).}
\begin{equation}\label{gauge matrix}
\raisebox{-0.4cm}{$\mathcal{G}=$}\begin{array}{cc}
\rule{0.5cm}{0pt}\overbrace{\rule{2.2cm}{0pt}}^{t_i^{(D)}} & \hspace{-0.6cm}\overbrace{\rule{3.8cm}{0pt}}^{t_i^{(I)}}\\
\multicolumn{2}{c}{ \left(
\begin{array}{cccccccc}
1      & 0      & \dots  & 0      & a_{11}   & a_{12}   & \dots  & a_{1N_I} \\
0      & 1      & \dots  & 0      & a_{21}   & a_{22}   & \dots  & a_{2N_I}\\
\vdots & \vdots & \ddots & \vdots & \vdots   & \vdots   & \ddots & \vdots \\
0      & 0      & \dots  & 1      & a_{N_D1} & a_{N_D2} & \dots  &
a_{N_DN_I}
\end{array}
\right)}
\end{array}.
\end{equation}
This matrix tells us how to divide the variables $t_i$ into $N_I$
independent variables $t_i^{(I)}$ and $N_D$ dependent variables
$t_i^{(D)}$. In the Schnabl gauge, it is convenient to use the
basis of $b$ and $c$ ghosts, in which the matrix $\mathcal{G}$ is
quite sparse and relatively easy to work with.

This form of $\mathcal{G}$ also immediately gives us a solution
for the dependent variables:
\begin{equation}\label{dependent variables solution}
t_i^{(D)}=-\sum_{j=1}^{N_I}a_{ij} t_i^{(I)}.
\end{equation}
This expression can be substituted into (\ref{EOM}), so that we
obtain equations only for the independent variables,
$f_i(t^{(I)},t^{(D)}(t^{(I)}))$.

However, these equations are still overdetermined, so in order to
solve them, we first have to select the projector $P$. In
principle, many choices are possible as long as the projector has
the correct rank $N_I$. For example, Kishimoto and Takahashi
\cite{KishimotoTakahashi1,KishimotoTakahashi2} used the Siegel
gauge projector in their calculations in the $a$-gauge. However,
there is one canonical choice for the projector.

The most natural choice for the projected equations is obtained by
substituting the dependent variables into the action, $S(t)\rar
S(t^{(I)},t^{(D)}(t^{(I)}))$, and by taking derivatives of this
restricted action with respect to the independent variables,
\begin{equation}\label{independent equations}
\frac{\del S(t^{(I)},t^{(D)}(t^{(I)}))}{\del t_i^{(I)}}=0.
\end{equation}
Using (\ref{action truncated}) and (\ref{dependent variables
solution}), we can derive an explicit formula for the canonical
projector $P_C$. It is closely related to the transpose of the
matrix $\mathcal{G}$:
\begin{equation}\label{gauge projector}
P_C=\left(
\begin{array}{cccccccc}
1         & 0         & \dots  & 0      & -a_{11}    & -a_{21}    & \dots  & -a_{N_D 1}   \\
0         & 1         & \dots  & 0      & -a_{12}    & -a_{22}    & \dots  & -a_{N_D 2}   \\
\vdots    & \vdots    & \ddots & \vdots & \vdots     & \vdots     & \ddots & \vdots       \\
0         & 0         & \dots  & 1      & -a_{1 N_I} & -a_{2 N_I} & \dots  & -a_{N_D N_I} \\
\end{array}
\right).
\end{equation}
If one decides to use the canonical projector, which we do in this
work, one can in principle avoid explicit use of the projector by
working directly with the restricted action as in
\cite{Arroyo:2017iis}. However, it is not possible to construct
the full matrix representation of the cubic vertex $V_{ijk}$ at
high levels due to large memory requirements, and we have to work
with the factorized matrices $V_{ijk}^{matter}$ and
$V_{ijk}^{ghost}$ only. The projector is not compatible with the
factorized vertices, therefore we have to apply it directly in
Newton's method.

When we work out Newton's method for the projected equations
(\ref{independent equations}), we find that the crucial step
changes to
\begin{equation}
\sum_j M_{ij}^{(P)}(t^{(n)})\Delta t_j^{(I)}=-f_i^{(P)}(t^{(n)}),
\end{equation}
where we define the projections of the Jacobian matrix and of the
equations of motion in terms of the non-projected quantities as
\begin{eqnarray}
M_{ij}^{(P)}&=&\sum_{k,l}P_{ik}P_{Cjl}M_{kl},\\
f_i^{(P)}&=&\sum_{j}P_{ij} f_j.
\end{eqnarray}
One of the steps of Newton's method allows us to find the change
of the independent variables, so the dependent variables can be
then easily computed using (\ref{dependent variables solution}).
Notice that, if one decides to use a non-canonical projector, the
Jacobian is multiplied by a different projector from each side.

As a matter of illustration, using a truncated level 4 string
field, we would like to explain in some detail how the method
above works. The string field up to level 4, following the
notation of Sen and Zwiebach \cite{Sen:1999nx}, is given by
\begin{eqnarray}\label{Psi lev 4}
|\Psi\ra&=&t c_1|0\ra+u c_{-1}|0\ra+v L^m_{-2}c_1|0\ra+w b_{-2}c_0 c_1 |0\ra+A L^m_{-4} c_1|0\ra+B L^m_{-2} L^m_{-2}c_1|0\ra \nonumber \\
&+&C c_{-3}|0\ra+D b_{-3}c_{-1}c_1|0\ra+E b_{-2}c_{-2}c_1|0\ra+F L^m_{-2}c_{-1}|0\ra+w_1 L^m_{-3}c_0|0\ra\nonumber \\
&+&w_2 b_{-2}c_{-1}c_0|0\ra+w_3 b_{-4}c_0 c_1|0\ra+w_4 L^m_{-2}
b_{-2}c_0 c_1|0\ra,
\end{eqnarray}
which means that the vector $t_i$ consists of the following
coefficients:
\begin{equation}\label{vector t}
t_i=(t,u,v,w,A,B,C,D,E,F,w_1,w_2,w_3,w_4).
\end{equation}
The Schnabl gauge condition $\mathcal{B}_0|\Psi\ra=0$ at this
level contains 5 independent equations
\begin{eqnarray}\label{gauge conditions}
w+\frac{2}{3} E&=&0, \\
w_i\!&=&0,\quad\quad\quad i=1,2,3,4. \nn
\end{eqnarray}
The matrix of gauge conditions $\mathcal{G}$ that follows from
(\ref{gauge conditions}) reads
\begin{equation}\label{G example}
\mathcal{G}=\left(
\begin{array}{cccccccccccccc}
 0 & 0 & 0 & 1 & 0 & 0 & 0 & 0 & \frac{2}{3} & 0 & 0 & 0 & 0 & 0 \\
 0 & 0 & 0 & 0 & 0 & 0 & 0 & 0 & 0           & 0 & 1 & 0 & 0 & 0 \\
 0 & 0 & 0 & 0 & 0 & 0 & 0 & 0 & 0           & 0 & 0 & 1 & 0 & 0 \\
 0 & 0 & 0 & 0 & 0 & 0 & 0 & 0 & 0           & 0 & 0 & 0 & 1 & 0 \\
 0 & 0 & 0 & 0 & 0 & 0 & 0 & 0 & 0           & 0 & 0 & 0 & 0 & 1 \\
\end{array}
\right).
\end{equation}
The matrix has reordered columns compared to (\ref{gauge matrix})
because we ordered the columns following (\ref{vector t}). We can
easily read off the dependent and independent variables,
\begin{eqnarray}
t_i^{(I)}&=&(t,u,v,A,B,C,D,E,F),\\
t_i^{(D)}&=&(w,w_1,w_2,w_3,w_4).\label{dependent variables}
\end{eqnarray}
Then we construct the canonical projector (\ref{gauge projector})
\begin{equation}\label{projector example}
P_C=\left(
\begin{array}{cccccccccccccc}
 1 & 0 & 0 & 0 & 0            & 0 & 0 & 0 & 0 & 0 & 0 & 0 & 0 & 0 \\
 0 & 1 & 0 & 0 & 0            & 0 & 0 & 0 & 0 & 0 & 0 & 0 & 0 & 0 \\
 0 & 0 & 1 & 0 & 0            & 0 & 0 & 0 & 0 & 0 & 0 & 0 & 0 & 0 \\
 0 & 0 & 0 & 1 & 0            & 0 & 0 & 0 & 0 & 0 & 0 & 0 & 0 & 0 \\
 0 & 0 & 0 & 0 & 0            & 1 & 0 & 0 & 0 & 0 & 0 & 0 & 0 & 0 \\
 0 & 0 & 0 & 0 & 0            & 0 & 1 & 0 & 0 & 0 & 0 & 0 & 0 & 0 \\
 0 & 0 & 0 & 0 & 0            & 0 & 0 & 1 & 0 & 0 & 0 & 0 & 0 & 0 \\
 0 & 0 & 0 & 0 & -\frac{2}{3} & 0 & 0 & 0 & 1 & 0 & 0 & 0 & 0 & 0 \\
 0 & 0 & 0 & 0 & 0            & 0 & 0 & 0 & 0 & 1 & 0 & 0 & 0 & 0 \\
\end{array}
\right).
\end{equation}
At this level, the projector has only one nontrivial element
$P_{85}=-\frac{2}{3}$. The percentage of nonzero elements is very
low even at higher levels, so we can work with the nontrivial
elements only and forget the rest of the matrix.

Unfortunately, we can not illustrate explicitly how the projector
acts on Newton's method because it would take far too much space,
but we will at least argue that it reproduces the correct
equations. So let us consider the expression
\begin{equation}
\sum_j P_{Cij}\frac{\del S(t)}{\del t_j}=0
\end{equation}
and compare it to (\ref{independent equations}). The action of the
projector at level 4 is mostly trivial. The projector reproduces
the original equations for variables $(t,u,v,A,B,C,D,F)$ and,
after substituting (\ref{gauge conditions}), we obtain the same
equations as in (\ref{independent equations}). The equations for
variables $(w_1,w_2,w_3,w_4)$ are correctly projected out, so the
equations for $E$ and $w$ are the only nontrivial check. The
projector mixes them together as
\begin{equation}
\frac{\del S}{\del E}-\frac{2}{3}\frac{\del S}{\del w}=0.
\end{equation}
It follows from (\ref{independent equations}) that
\begin{eqnarray}
0&=&\frac{\del S(E,w(E),\dots)}{\del E} \nonumber \\
&=&\frac{\del S}{\del E}+\frac{\del w(E)}{\del E}\frac{\del S}{\del w} \\
&=&\frac{\del S}{\del E}-\frac{2}{3}\frac{\del S}{\del w},
\nonumber
\end{eqnarray}
from which we observe that the projector reproduces the correct
equation.

For consistency, we have checked up to level 10 that this method
provides the same solution for tachyon vacuum in the Schnabl gauge
as the approach used in \cite{Arroyo:2017iis}. Furthermore, we
improved the results from \cite{Arroyo:2017iis} to level 24.

\section{Coefficients of the tachyon condensate}\label{sec:coefficients}

Using $(L,3L)$ level truncation computations, we have determined
the tachyon condensate in the Schnabl gauge up to level $L=24$.
For practical purposes, we cannot provide the complete list of all
the coefficients of the tachyon condensate up to this level 24
(there are 54678 coefficients). Instead, we show in table
\ref{tab:coefficients} the tachyon coefficient together with some
other low level coefficients. We would like to compare these
coefficients with the analytic solution from
\cite{Schnabl:2005gv}. To do so, we will predict the asymptotic
values of these coefficients using the methods studied in appendix
\ref{sec:extrapolations}.

For a matter of illustration, we will explain in some detail the
analysis of the data for the tachyon coefficient $t$. For the rest
of the coefficients, we will only provide the results.

Let us start with the extrapolation of the tachyon coefficient
data by means of functions in $1/L$ of the form (\ref{1overLI})
\begin{align}\label{1overLtach}
T^{(L_{min},L_{max})}_M(L)=a_0 + \sum_{n=1}^{M} \frac{a_n}{L^n},
\end{align}
where we are using the notation introduced in appendix
\ref{sec:extrapolations}. Assuming that we have data from level
$L_{min}$ to $L_{max}$, the order $M$ can be chosen between 1 and
$(L_{max}-L_{min})/2$. For reasons explained in appendix
\ref{sec:extrapolations}, we will use mainly polynomials of
maximal orders $N=(L_{max}-L_{min})/2$.

The antepenultimate row and the second column of table
\ref{tab:coefficients} shows the asymptotic value of the tachyon
coefficient $T^{(4,24)}_{10}(\infty) \approx 0.5457$, which has
been obtained using data from levels $(4,24)$ and maximal order
$N=10$ fit\footnote{In this section, we have decided to remove
level 2 data from the analysis because they lead to lesser
stability of $N\rar\inf$ extrapolations. The most likely reason is
that the solution at this level is the same as in the Siegel
gauge.}. We find a rough agreement between this asymptotic value
and the analytical value $t_{analytic}=0.55346558$, however, the
precision is not sufficiently good. Let us see if we can improve
the precision. The second column in table \ref{tab:coefN1} shows
the asymptotic values $\lim_{L\rightarrow
\infty}T^{(4,2N+4)}_N(L)$ obtained by means of maximal order
extrapolations with varying number of data points. We observe that
the results change nontrivially as we add more points and the
general trend is that they approach the analytical value with
increasing $N$. Therefore is worth trying to make a second,
$N\rar\inf$, extrapolation following the procedure described in
appendix \ref{sec:extrapolations}.

\begin{table}[t]
\centering
\begin{tabular}{|c|lllll|}\hline
$L$         & $t$         & $u$         & $v$         & $\ps w$ &
$\ps w_1$          \\\hline
2           & 0.544204232 & 0.190190387 & 0.055963664 & $\ps 0          $ & $\ps 0           $ \\
4           & 0.548938521 & 0.242291544 & 0.069463998 & $   -0.022500108$ & $\ps 0           $ \\
6           & 0.548315148 & 0.271907177 & 0.077794757 & $   -0.036964272$ & $   -0.0001097937$ \\
8           & 0.547321883 & 0.291164859 & 0.083458462 & $   -0.046949656$ & $   -0.0001568187$ \\
10          & 0.546508411 & 0.304819814 & 0.087592014 & $   -0.054296587$ & $   -0.0001789026$ \\
12          & 0.545894444 & 0.315095030 & 0.090767237 & $   -0.059969107$ & $   -0.0001895219$ \\
14          & 0.545435158 & 0.323166082 & 0.093300578 & $   -0.064511081$ & $   -0.0001943355$ \\
16          & 0.545089161 & 0.329713368 & 0.095381100 & $   -0.068251134$ & $   -0.0001960127$ \\
18          & 0.544825972 & 0.335158842 & 0.097128893 & $   -0.071399574$ & $   -0.0001959212$ \\
20          & 0.544624075 & 0.339778735 & 0.098624073 & $   -0.074097466$ & $   -0.0001948079$ \\
22          & 0.544468282 & 0.343761952 & 0.099922283 & $   -0.076443182$ & $   -0.0001931002$ \\
24          & 0.544347732 & 0.347242371 & 0.101063472 & $
-0.078507578$ & $   -0.0001910522$ \\\hline $L\rar\inf$ & 0.54572
& 0.409       & 0.122       & $   -0.117      $ & $   -0.0000910 $
\\\hline $N\rar\inf$ & 0.55340     & 0.431       & 0.130       & $
-0.132      $ & $\ps 0.0000033   $ \\\hline Analytic    &
0.55346558  & 0.45661043  & 0.13764616  & $   -0.14421001 $ & $\ps
0           $ \\\hline
\end{tabular}
\caption{Some coefficients of the numerical solution in the
Schnabl gauge up to level 24. We show coefficients of the level 2
string field plus the coefficient $w_1$ from level 4. These
coefficients are denoted following (\ref{Psi lev 4}). The last
three rows show the results of: $L\rar \inf$ extrapolations (using
level $(4,24)$ data and order $N=10$), $N\rar \inf$ extrapolations
(using order $M=5$), and the expected analytical values
respectively.} \label{tab:coefficients}
\end{table}

\begin{table}[t]
\centering
\begin{tabular}{|c|lllll|}\hline
$N$ & $t$      & $u$      & $v$      & $\ps w$     & $\ps w_1$
\\\hline
1   & 0.547068 & 0.331138 & 0.094456 & $-0.065893$ & $-0.000109794$ \\
2   & 0.541616 & 0.366737 & 0.106443 & $-0.087919$ & $-0.000297894$ \\
3   & 0.541628 & 0.380829 & 0.111773 & $-0.097807$ & $-0.000221255$ \\
4   & 0.542646 & 0.388399 & 0.114684 & $-0.103198$ & $-0.000165502$ \\
5   & 0.543345 & 0.393854 & 0.116718 & $-0.106952$ & $-0.000138074$ \\
6   & 0.543882 & 0.398126 & 0.118282 & $-0.109836$ & $-0.000122554$ \\
7   & 0.544389 & 0.401572 & 0.119533 & $-0.112142$ & $-0.000110590$ \\
8   & 0.544876 & 0.404346 & 0.120537 & $-0.113988$ & $-0.000102323$ \\
9   & 0.545315 & 0.406788 & 0.121407 & $-0.115594$ & $-0.000091993$ \\
10  & 0.545722 & 0.408636 & 0.122077 & $-0.116808$ &
$-0.000090974$ \\\hline
\end{tabular}
\caption{Asymptotic values of the tachyon coefficient and the
other coefficients from table \ref{tab:coefficients}. The results
were obtained using functions $T^{(4,2N+4)}_N(L)$ (and the
analogues for the other coefficients) with $N$ varying from 1 to
10.} \label{tab:coefN1}
\end{table}

By extrapolating the ten data points given in table
\ref{tab:coefN1}, we can get a result that is much closer to the
expected analytical one. Let us see explicitly how this works. To
extrapolate the data in table \ref{tab:coefN1}, we employ the
following function:
\begin{align}
\label{tRN} t_M(N)=r_0 + \sum_{n=1}^{M} \frac{ r_n}{ N^n},
\end{align}
where the coefficients $r_0,r_1,\cdots,r_M$ can be determined
using the least squares fitting technique. In principle, the value
of the fit degree $M$ could be chosen from 1 and 9. However, the
values in table \ref{tab:coefN1} do not lie on a smooth curve like
the original data. Therefore, for high values of $M$, it turns out
that the coefficients $r_0,r_1,\cdots,r_M$ have large values and
the extrapolations are unstable.

In the second column of table \ref{tab:coefinf}, we show the
results of these extrapolations for the tachyon coefficient
$\lim_{N \rightarrow \infty}t_M(N)$ for $2\leq M\leq 7$. The
results are not too far from each other, which means that for
lower $M$, this extrapolation procedure is relatively stable. By
repeating the same analysis for other coefficients, we have
reached a conclusion that $M=5$ is the best overall choice, so
these results are shown in the penultimate row of table
\ref{tab:coefficients} denoted by $N \rar \infty$. For the tachyon
coefficient, the result of the $M=5$ extrapolation agrees very
well with the analytical value. The precision is about $0.01\%$,
but, considering that the other values in table \ref{tab:coefinf}
are further away, such good precision is probably coincidental.

When it comes to other coefficients, we decided to show the
coefficients from level 2 string field, $u$, $v$ and $w$, and the
coefficient $w_1$ from level 4, which is equal to zero for the
analytic solution. We analyzed them in the same way as the tachyon
coefficient, the results are shown in tables
\ref{tab:coefficients}, \ref{tab:coefN1} and \ref{tab:coefinf} in
columns three to six. We can conclude that these coefficients
behave similarly as the tachyon coefficient. The level infinity
extrapolations roughly agree with the analytic values, but the
precision is low. Therefore we again performed the $N \rar \inf$
extrapolations, whose results are much closed to the expected
analytic values, although the precision is lower than in the case
of the tachyon coefficient.

\begin{table}[!]
\centering
\begin{tabular}{|c|lllll|}\hline
$M$      & $t$      & $u$      & $v$      & $\ps w$     & $\ps
w_1$         \\\hline
2        & 0.547650 & 0.419700 & 0.126421 & $-0.124876$ & $\ps 0.000020124$ \\
3        & 0.549081 & 0.425743 & 0.128172 & $-0.128001$ & $   -0.000052372$ \\
4        & 0.549252 & 0.432140 & 0.130259 & $-0.131814$ & $   -0.000073693$ \\
5        & 0.553396 & 0.431237 & 0.130247 & $-0.131695$ & $\ps 0.000003298$ \\
6        & 0.553639 & 0.428358 & 0.129141 & $-0.129134$ & $   -0.000121455$ \\
7        & 0.548023 & 0.416816 & 0.125705 & $-0.120888$ & $
-0.000821313$ \\\hline Analytic & 0.553466 & 0.456610 & 0.137646 &
$-0.144210$ & $\ps 0          $ \\\hline
\end{tabular}
\caption{The results of $N\rar\inf$ extrapolations of the data
from table \ref{tab:coefN1} using $1/N$ polynomials of order $M$.}
\label{tab:coefinf}
\end{table}

Finally, we would like to comment more about high level behavior
of the tachyon coefficient $t$. In the reference
\cite{Arroyo:2017iis}, using results up to level $L=10$, the
authors predicted that the tachyon coefficient reaches a local
minimum value at level $L \sim 26$, to then turn back to approach
the expected analytical result as $L \rightarrow \infty$. Using
the additional data up to level 24, we will show that this local
minimum probably exists, but at a much higher level.

\begin{figure}[h]
\centering
\includegraphics[width=6.0in,height=90mm]{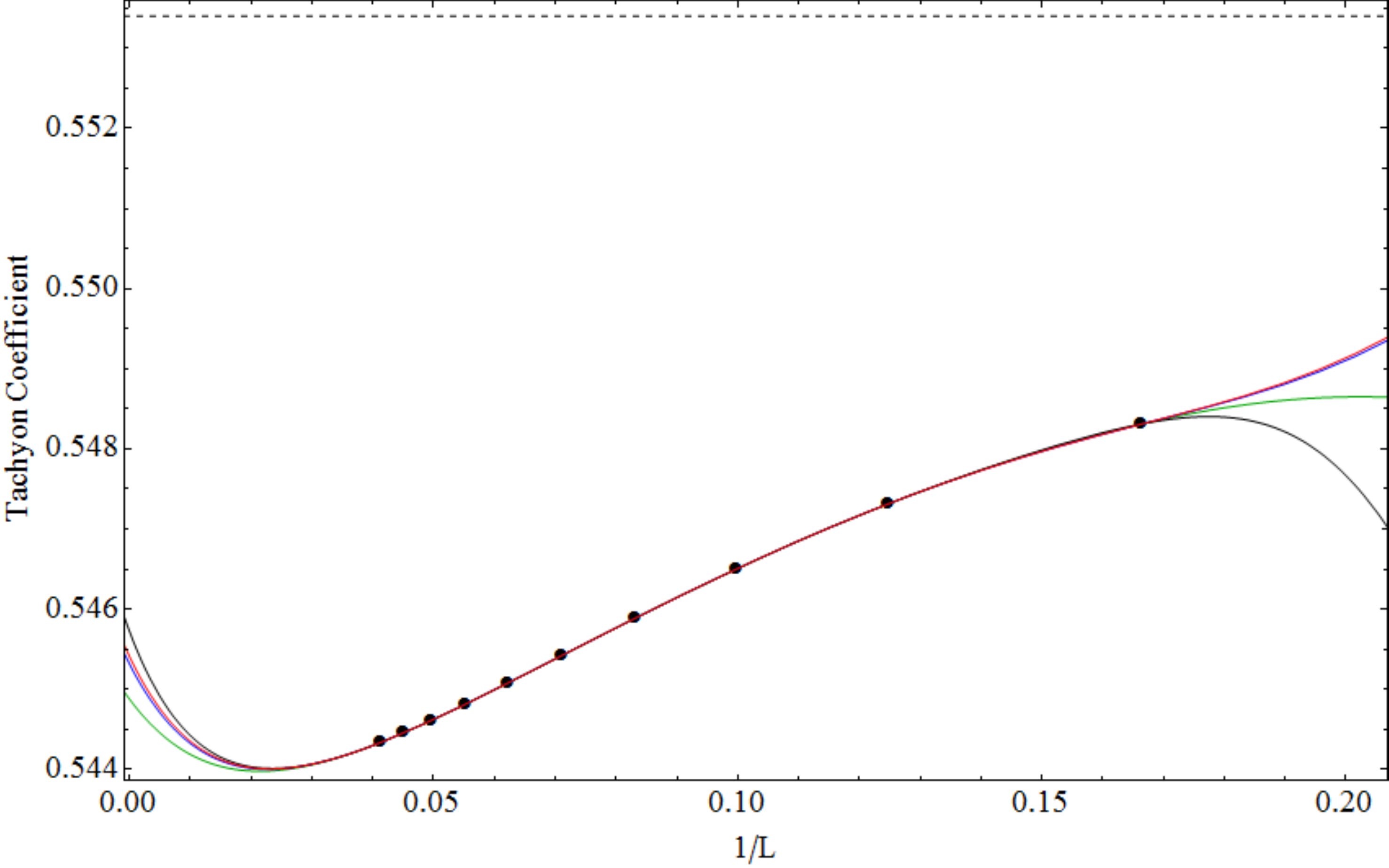}
\caption{The tachyon coefficient data up to level $L=24$ in the
Schnabl gauge as a function of $1/L$. The green solid curve
represents the function $T^{(4,20)}_8(L)$. The blue solid curve
represents the function $T^{(4,22)}_9(L)$. The black solid curve
represents the function $T^{(4,24)}_{10}(L)$. The red solid curve
represents the function $T^{(2,24)}_{11}(L)$. The dashed line
represents the analytical value $0.553465$.} \label{fig:tachyon}
\end{figure}

We can estimate at what level this local minimum arises using the
functions $T^{(4,20)}_8(L)$, $T^{(4,22)}_9(L)$ and
$T^{(4,24)}_{10}(L)$. In figure \ref{fig:tachyon}, we plot these
functions together with the tachyon coefficient data. And for
reference, we also plot the function $T^{(2,24)}_{11}(L)$ that
uses the maximal amount of data points. We have computed the local
minimum of $T^{(4,20)}_8(L)$, $T^{(4,22)}_9(L)$ and
$T^{(4,24)}_{10}(L)$ and obtained the corresponding values $L \sim
46$, $L \sim 43$ and $L \sim 42$ respectively.

Although these estimates of the local minimum are not very
reliable, from the trend observed in the data up to level 24 (see
figure \ref{fig:tachyon}), we can state with certainty that this
minimum does not happen at $L=26$ as claimed in
\cite{Arroyo:2017iis}.

\section{The vacuum energy and the gauge invariant overlap} \label{sec:Energy}
The values for the vacuum energy obtained by direct $(L,3L)$ level
truncation computations in the Schnabl gauge up to level $L=24$,
are given in table \ref{tab:Energy1}. For purposes of comparison,
we have also written the data for the vacuum energy in the Siegel
gauge up to level $L=30$ \cite{KudrnaUniversal}.

\begin{table}[t]
\centering
\begin{tabular}{|c|l|l|}\hline
$L$ & $\ps E^{Sch}$     & $\ps E^{Sie}$     \\\hline
2   & $-0.959376599521$ & $-0.959376599521$ \\
4   & $-0.994651904750$ & $-0.987821756244$ \\
6   & $-1.003983765388$ & $-0.995177120537$ \\
8   & $-1.007110280219$ & $-0.997930183378$ \\
10  & $-1.008189759705$ & $-0.999182458475$ \\
12  & $-1.008466266815$ & $-0.999822263312$ \\
14  & $-1.008396790194$ & $-1.000173729946$ \\
16  & $-1.008173012946$ & $-1.000375451894$ \\
18  & $-1.007882751544$ & $-1.000493711466$ \\
20  & $-1.007568595810$ & $-1.000562954585$ \\
22  & $-1.007251843369$ & $-1.000602262320$ \\
24  & $-1.006943179985$ & $-1.000622749436$ \\
26  &                   & $-1.000631156455$ \\
28  &                   & $-1.000631706784$ \\
30  &                   & $-1.000627117967$ \\\hline $\inf$ &
$-0.99949$     & $-1.000009$       \\\hline
\end{tabular}
\caption{$(L, 3L)$ level-truncation results for the vacuum energy
in the Schnabl gauge $E^{Sch}$ as well as in the Siegel gauge
$E^{Sie}$.} \label{tab:Energy1}
\end{table}

In the reference \cite{Arroyo:2017iis}, the energy in the Schnabl
gauge up to level $L=10$ was computed using direct $(L,3L)$ level
truncation computations and, by extrapolating this level $L=10$
data, it was predicted that the vacuum energy reaches a local
minimum at level $L \sim 12$, to then turn back to approach its
expected analytical value $-1$ \footnote{The energy in this
section is normalized according to the convention adopted in
reference \cite{Arroyo:2017iis}. Note that, to match the
convention adopted in this paper with the convention of
\cite{KudrnaUniversal}, the value of the energy should be shifted
by 1.} asymptotically as $L \rightarrow \infty$. Note that the
existence of this local minimum can be confirmed by looking
directly at the data given in table \ref{tab:Energy1}.

Since now we have data for the energy up to level $L=24$, it would
be nice to see what the behavior of the energy at higher levels
is. In order to extrapolate our data, we use the procedure
introduced in appendix \ref{sec:extrapolations}. We consider
polynomial functions of the form (\ref{1overLI}), where we choose
the maximal order $N=(L_{max}-L_{min})/2$.

\begin{figure}[h]
\centering
\includegraphics[width=6.0in,height=90mm]{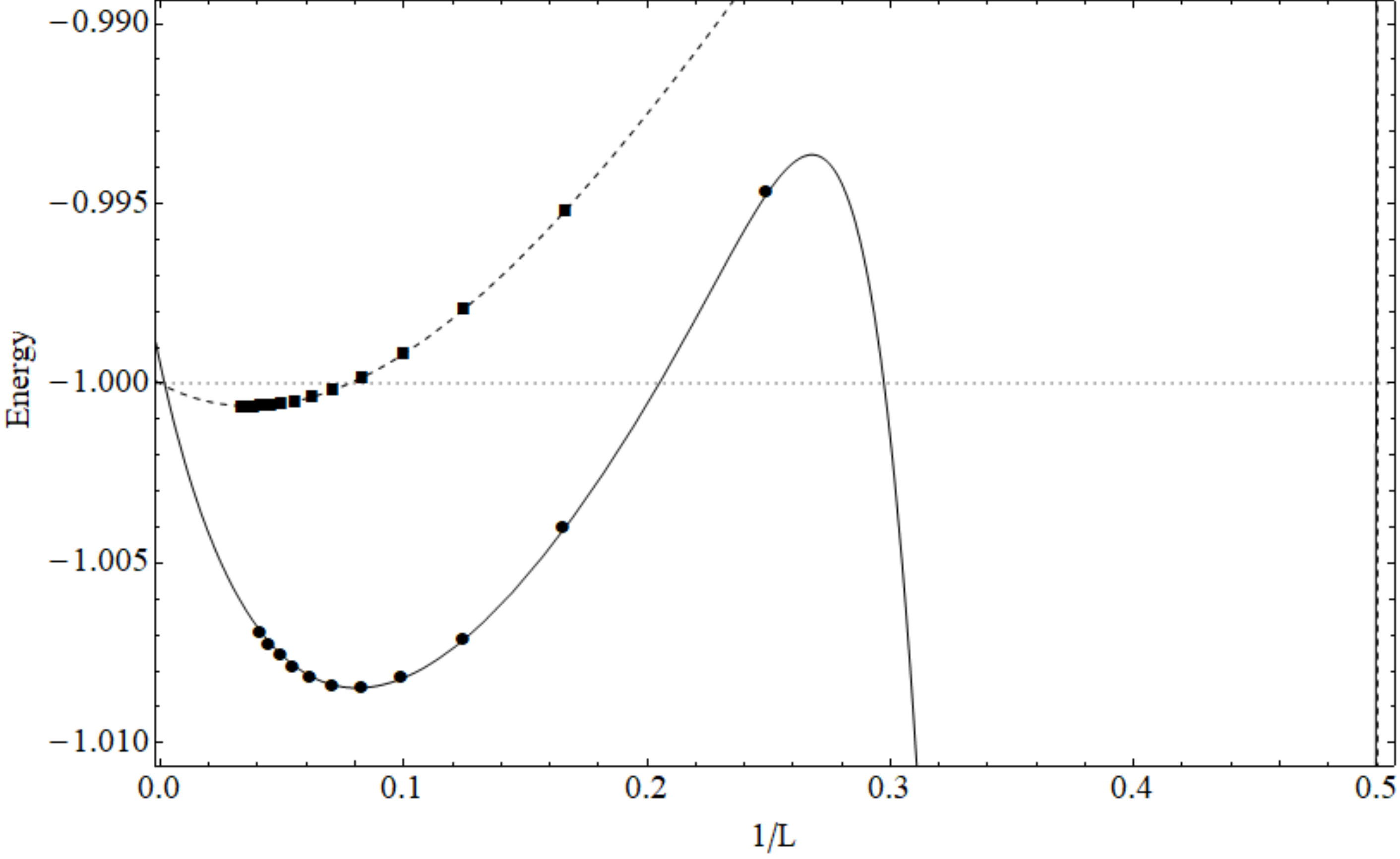}
\caption{Data for the energy in the Schnabl gauge as well as in
the Siegel gauge as a function of $1/L$. The solid curve
represents the function $E^{(2,24)}_{11}$ that interpolates the
level $L=24$ data in the Schnabl gauge. The dashed curve
represents the function $E^{(2,30)}_{14}$ that interpolates the
level $L=30$ data in the Siegel gauge. The dotted line represents
the analytical value $-1$.} \label{fig:SiegelSchnablEnergy}
\end{figure}

Employing the data given in table \ref{tab:Energy1}, we can
construct  functions $E^{(2,2N+2)}_N(L)$ up to order $N=11$ in the
Schnabl gauge, and up to order $N=14$ in the Siegel gauge. The
asymptotic values $\lim_{L \rightarrow \infty} E^{(2,2N+2)}_N(L)$
in both gauges are provided in table \ref{tab:EnergyN}. In figure
\ref{fig:SiegelSchnablEnergy}, we show the plot of
$E^{(2,24)}_{11}$ in the Schnabl gauge and of $E^{(2,30)}_{14}$ in
the Siegel gauge. We observe that the precision of the fit in the
Siegel gauge is much better. Even though the function
$E^{(2,24)}_{11}$ in the Schnabl gauge fits the corresponding data
well, we have noted that, at a level close to $L\sim 590$, the
value of the extrapolated energy overshoots the expected result
$-1$ and the asymptotic value when the level goes to infinity is
close to $-0.9995$. Other functions $E^{(2,2N+2)}_N(L)$ with high
$N$ behave similarly.

We note that the size of this overshooting is similar to the
dispersion of the results, and therefore we can say that it lies
within the error of the extrapolation procedure. Have a closer
look at table \ref{tab:EnergyN}. As in the previous section, there
is a problem with stability of the extrapolations. We observe that
the differences between these asymptotic values are much higher in
the Schnabl gauge than in the Siegel gauge, and that these changes
are not random, but they follow a certain trend. Similar behavior
can be observed if we make other variations in the extrapolation
procedure, for example, we can keep $L_{max}=24$ and vary
$L_{min}$ or we can use less than maximal order extrapolations.

The simplest conclusion is that, based on the differences in table
\ref{tab:EnergyN}, the asymptotic value of the energy in the
Schnabl gauge has an error of order $10^{-4}$ and therefore the
energy agrees with the analytical value within this error.

We have tried to make a second, $N\rar\inf$ extrapolation of the
values in table \ref{tab:EnergyN} similarly to section
\ref{sec:coefficients}, but the results do not lead to a better
precision. Diverse extrapolations in $1/N$ agree on a maximum
around $N=10$, which means that these functions start to approach
$-1$ again above this value, but the asymptotic values of these
functions have too large dispersion to be trustworthy.

\begin{table}[!t]
\centering
\begin{tabular}{|c|l|l|}\hline
$N$ & $\ps E^{Sch}$  & $\ps E^{Sie}$     \\\hline
1   & $-1.029927$ & $-1.0162669$ \\
2   & $-1.019008$ & $-1.0066983$ \\
3   & $-1.007440$ & $-1.0010884$ \\
4   & $-1.003143$ & $-1.0001414$ \\
5   & $-1.001520$ & $-1.0000732$ \\
6   & $-1.000607$ & $-1.0000552$ \\
7   & $-1.000045$ & $-1.0000371$ \\
8   & $-0.999722$ & $-1.0000256$ \\
9   & $-0.999568$ & $-1.0000178$ \\
10  & $-0.999476$ & $-1.0000137$ \\
11  & $-0.999486$ & $-1.0000131$ \\
12  &             & $-1.0000096$ \\
13  &             & $-1.0000094$ \\
14  &             & $-1.0000091$ \\\hline
\end{tabular}
\caption{Extrapolations of the energy using order $N$ polynomials
and level $(2,2N+2)$ data in the Schnabl and in the Siegel gauge.}
\label{tab:EnergyN}
\end{table}

Another quantity analyzed is the so-called gauge invariant overlap
or Ellwood's invariant. For a given solution $\Psi$ of the string
field equation of motion, let us define this gauge invariant
quantity $E_0$ as\footnote{We are using the same convention for
$E_0$ as given in reference \cite{KudrnaUniversal}, which is
shifted by $-1$ when compared with the convention of references
\cite{Hashimoto:2001sm,Ellwood:2008jh}. The invariant can be
easily evaluated using the conservation laws in
\cite{Kudrna:2012re}.}
\begin{align}\label{inv1}
E_0  = 1+\langle \mathcal{V} |\Psi\rangle =1+\langle
I|\mathcal{V}(i) |\Psi\rangle,
\end{align}
where $| I \rangle$ is the identity string field and
$\mathcal{V}(i)$ is an on-shell closed string vertex operator
$\mathcal{V}=c \tilde c V^{\text{m}}$, which is inserted at the
midpoint of the string field $\Psi$. Using the $(L,3L)$ level
truncated numerical solution in the Schnabl gauge, we computed
$E_0$ up to level $L=24$. The results are shown in table
\ref{tab:E0}. Note that $E_0$ appears to approach the expected
analytical value $E_0 = 0$.

\begin{table}[!tb]
\centering
\begin{tabular}{|c|l|}\hline
$L$ & $\ps E_0$   \\\hline
2   & $\ps 0.110138182891$ \\
4   & $\ps 0.068341344418$ \\
6   & $\ps 0.047847933137$ \\
8   & $\ps 0.034175803769$ \\
10  & $\ps 0.027266149826$ \\
12  & $\ps 0.020987152232$ \\
14  & $\ps 0.017848318168$ \\
16  & $\ps 0.014361780518$ \\
18  & $\ps 0.012688927165$ \\
20  & $\ps 0.010510916529$ \\
22  & $\ps 0.009525064795$ \\
24  & $\ps 0.008051965768$ \\\hline $\infty$ & $   -0.0015$
\\\hline
\end{tabular}
\caption{$(L, 3L)$ level-truncation results for the gauge
invariant overlap $E_0$ in the Schnabl gauge.} \label{tab:E0}
\end{table}

\begin{figure}[h]
\centering
\includegraphics[width=6.0in,height=90mm]{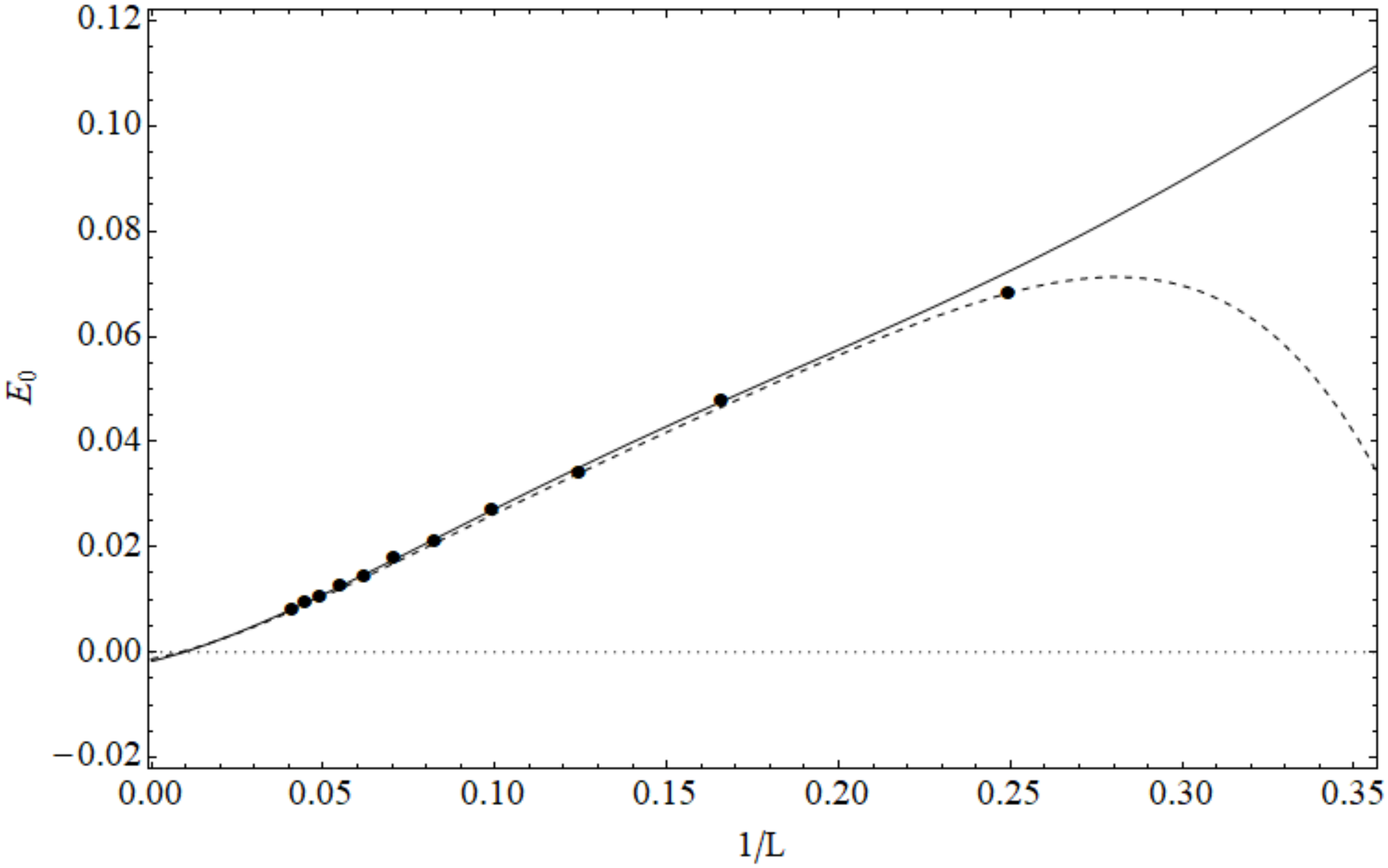}
\caption{Data for the gauge invariant quantity $E_0$ up to level
$L=24$ in the Schnabl gauge as a function of $1/L$. The solid
curve represents the function $f_5(L)$, while the dashed curve
represents the function $g_5(L)$. The dotted line represents the
expected analytical value $E_0=0$.} \label{fig:E01overL}
\end{figure}

Let us predict the value for this gauge invariant quantity $E_0$
when $L \rightarrow \infty$. To do this, we need to extrapolate
the corresponding data given in table \ref{tab:E0}. If we use the
$1/L$ polynomial interpolation given by equation (\ref{1overLI})
and the data of table \ref{tab:E0}, we can obtain an order $N=11$
extrapolation. However when we compute the asymptotic value $L
\rightarrow \infty$ of this order $N=11$ function, we get
$3.3629$, which is clearly far away from the expected value
$E_0=0$. So, we observe that for the gauge invariant overlap, we
have to extrapolate levels $4k+2$ and $4k$ separately\footnote{See
\cite{KudrnaThesis} for more detailed discussion of extrapolations
of Ellwood invariants.}. Namely, we divide the twelve data points
of table \ref{tab:E0} into two sets of six data points, the ones
at levels $L=(2,6,10,14,18,22)$ and $L=(4,8,12,16,20,24)$ and for
each set of data points, we use a $1/L$ polynomial function of
order $N=5$. For instance, using the data of levels $L=4k+2$,
where $k=0,1,\cdots,5$, we obtain the following result
\begin{align}
\label{E051} f_5(L) = -0.00172+ \frac{0.159869}{L}+
\frac{2.64626}{L^2} - \frac{18.1245}{L^3} + \frac{51.6933}{L^4} -
\frac{51.0373}{L^5}.
\end{align}
Note that when $L \rightarrow \infty$, the above expression
approaches $-0.00172$, which is close to the expected analytical
value of the gauge invariant overlap, $E_0 = 0$. Now, if we use
the data of levels $L=4k$, where $k=1,\cdots,6$, we obtain
\begin{align}
\label{E052} g_5(L) = -0.00129+ \frac{0.129984}{L}+
\frac{3.13195}{L^2} - \frac{24.3127}{L^3} + \frac{88.936}{L^4} -
\frac{129.155}{L^5}.
\end{align}
So in this case, the asymptotic value turns out to be $-0.00129$,
which is also close to the expected value of $E_0=0$. By taking
the average of these two asymptotic values, we get $-0.0015$,
which means that this extrapolation technique gives us the
invariant $E_0$ with a relative precision of $0.15\%$.

For illustrative purposes, we have plotted in figure
\ref{fig:E01overL} the data of the gauge invariant quantity given
in table \ref{tab:E0} together with the functions $f_5(L)$ and
$g_5(L)$.

\section{Checking some symmetries}\label{sec:identities}

In reference \cite{Schnabl:2005gv}, the author showed that the
analytical solution in the Schnabl gauge satisfies
$K^{matter}_1\Psi=0$ and $[K^{matter}_1,Q]\Psi=0$, where
$K^{matter}_1 = L^{matter}_{1}+L^{matter}_{-1}$. A consequence of
this symmetry is as follows: Consider the expansion of the
analytical solution $\Psi$ in the Virasoro basis of $L_0$
eigenstates, for instance, this expansion up to level $L=4$ is
given in equation (\ref{Psi lev 4}), where the exact coefficients
are shown in table \ref{tab:coefanalytic}. Employing the
symmetries $K^{matter}_1\Psi=0$ and $[K^{matter}_1,Q]\Psi=0$, it
is possible to show that the exact coefficients satisfy the
following relations
\begin{align}
\label{ww1} w_1 = 0 , \\
\label{d1} d_1 \equiv 5A + 3B + v = 0 , \\
\label{d2} d_2 \equiv 20A + 12B + 4D - 4F - 8w_1 = 0 , \\
\label{d4} d_3 \equiv 15A + 9B + v + w - 10w_1 + 5w_3 + 3w_4 = 0 .
\end{align}

Using the values of the coefficients given in table
\ref{tab:coefanalytic}, we can explicitly verify that these
coefficients satisfy the above identities. There is one more
independent identity
\begin{align}
\label{d4} d_4 \equiv 2A + 4D - 3E + 2F - 3w_2 + 3w_4 = 0.
\end{align}
In reference \cite{Schnabl:2005gv}, the author failed in finding a
simple origin for this identity, it might be just an accidental
symmetry.

\begin{table}\nonumber
\centering
\begin{tabular}{|llll|}\hline
$t=0.55346558$  & $\;\;\; A=-0.03027758$ & $\;\;\; E=0.17942652$ & $\;\;\; w_1=0$ \\
$u=0.45661043$  & $\;\;\; B=0.00458058$  & $\;\;\; F=0.02274827$ & $\;\;\; w_2=0.02094354$ \\
$v=0.13764616$  & $\;\;\; C=-0.16494614$ &                       & $\;\;\; w_3=0.08898226$ \\
$w=-0.14421001$ & $\;\;\; D=0.16039444$  &                       &
$\;\;\; w_4=-0.00846965$ \\\hline
\end{tabular}
\caption{Coefficients of Schnabl analytical solution up to level
4.} \label{tab:coefanalytic}
\end{table}

\begin{table}[t]
\centering
\begin{tabular}{|c|llll|}\hline
$L$         & $d_1$        & $\ps d_2$          & $\ps d_3$ & $\ps
d_4$          \\\hline
4           & 0.0418100836 & $   -0.0051417488$ & $   -0.0359978557$ & $   -0.0313612657$ \\
6           & 0.0389000577 & $\ps 0.0004040458$ & $   -0.0326194412$ & $   -0.0311201802$ \\
8           & 0.0368605116 & $\ps 0.0028520778$ & $   -0.0305285804$ & $   -0.0300137406$ \\
10          & 0.0352699472 & $\ps 0.0041306703$ & $   -0.0290140111$ & $   -0.0286764813$ \\
12          & 0.0339700353 & $\ps 0.0048583975$ & $   -0.0278298640$ & $   -0.0273399210$ \\
14          & 0.0328764762 & $\ps 0.0052915301$ & $   -0.0268613392$ & $   -0.0260829188$ \\
16          & 0.0319371821 & $\ps 0.0055535049$ & $   -0.0260450246$ & $   -0.0249272454$ \\
18          & 0.0311173345 & $\ps 0.0057100850$ & $   -0.0253419397$ & $   -0.0238732195$ \\
20          & 0.0303924516 & $\ps 0.0057989507$ & $   -0.0247263128$ & $   -0.0229134689$ \\
22          & 0.0297446764 & $\ps 0.0058429156$ & $   -0.0241801873$ & $   -0.0220384197$ \\
24          & 0.0291606055 & $\ps 0.0058564022$ & $ -0.0236905443$
& $   -0.0212384646$ \\\hline $L\rar\inf$ & 0.0158 & $\ps 0.00256
$ & $   -0.0127      $ & $   -0.0031      $
\\\hline $N\rar\inf$ & 0.0074       & $   -0.00082     $ & $
-0.0059      $ & $\ps 0.0059      $ \\\hline Analytic    & 0 &
$\ps 0           $ & $\ps 0           $ & $\ps 0           $
\\\hline
\end{tabular}
\caption{Results for the identities (\ref{d1})-(\ref{d4}) obtained
from the numerical solution in the Schnabl gauge up to level 24.}
\label{tab:identityds}
\end{table}

We would like to verify whether the numerical solution satisfies
the identities (\ref{ww1})-(\ref{d4}). Actually, the identity
$w_1=0$ appears to be satisfied by the numerical solution when $L
\rightarrow \infty$, see the last column of table
\ref{tab:coefficients}.

Using numerical results for the values of the coefficients that
have been obtained by means of $(L,3L)$ level truncation
computations up to level $L=24$ for the remaining identities
(\ref{d1})-(\ref{d4}), we found the results shown in table
\ref{tab:identityds}.

Since these identities are linear combinations of string field
coefficients, they can be extrapolated using the methods of
appendix \ref{sec:extrapolations}. The line denoted by $L \rar
\infty$ shows results obtained by means of functions of the form
\ref{1overLI} (in the case of the identities
(\ref{d1})-(\ref{d4}), we denote these functions as
$d^{(4,24)}_{i,10}(L)$, $i=1,2,3,4$). The values shown in the line
with $N \rar \infty$ were obtained by fits of the form given in
equation (\ref{1overNI1}) with degree $M=5$. With the exception of
the last identity $d_4$, the $N \rar \infty$ extrapolations
improve the results of the usual $L \rar \infty$ extrapolations.

From the results given in table \ref{tab:identityds}, we can
conclude that the identities (\ref{d1})-(\ref{d4}) are satisfied.

\section{Out-of-gauge equations of motion}\label{sec:out-of-gauge}

In the level truncation approach to SFT, it is possible to solve
only a subset of the full equations of motion (\ref{equations
solved}). However, in order to verify that the tachyon vacuum is a
physical solution, we should check whether it violates the
remaining equations (\ref{equations unsolved}). In the ideal case,
the violation of these equations approaches zero as the level goes
to infinity.

The evaluation of these out-of-gauge equations in the Schnabl
gauge is much simpler than in the Siegel gauge because we have
access to the full set of cubic vertices. Therefore we decided to
check the equations that come from derivations of the action with
respect to the dependable variables (\ref{dependent variables}).
The results are summarized in table \ref{tab:equations}, including
extrapolations to infinite level.

We extrapolate the data using functions of the form
(\ref{1overLI}), where we choose $L_{min}=4$ and the maximal
possible order $N=10$. We omit the data points at level 2 because
they do not follow the trend given by the remaining data points.
There are probably two reasons for that. The solution at level 2
is the same as in the Siegel gauge and the equations for level 4
fields cannot be satisfied well while we are still at level 2. We
have also tried to do $N\rar \inf$ extrapolations like in section
\ref{sec:coefficients}, but when we computed the analogue of table
\ref{tab:coefN1}, we have found that these results are not orderly
enough to allow meaningful extrapolations in $1/N$.

The first equation $f_w$ is a direct analogue of $\Delta_S$ in
\cite{KudrnaUniversal}. We can see that it quickly decreases with
the level, although somewhat slower than in case of the Siegel
gauge solution. The other equations are also satisfied well and
most of them monotonously approach zero (excluding the exceptional
level 2). The extrapolations improve the values at level 24 by
another order, so it is most likely that these equations are
asymptotically satisfied. The only exception to this trend is the
equation $f_{w_4}$, which overshoots zero. However, its
extrapolation predicts that it has a maximum around level 37 to
then turn back and go to zero as well.

\begin{table}\nonumber
\centering
\begin{tabular}{|c|lllll|}\hline
$L$      & $\ps f_w$        & $\ps f_{w_1}$   & $f_{w_2}$  & $\ps
f_{w_3}$    & $\ps f_{w_4}$    \\\hline
2        & $\ps 0.03332993$ & $   -0.0748098$ & 0.04390940 & $\ps 0.14787107$ & $\ps 0.04589104$ \\
4        & $\ps 0.01648496$ & $   -0.1755207$ & 0.00810403 & $   -0.03525052$ & $   -0.02910861$ \\
6        & $\ps 0.01026732$ & $   -0.1129462$ & 0.00682673 & $   -0.02057664$ & $   -0.01258739$ \\
8        & $\ps 0.00719823$ & $   -0.0830693$ & 0.00595911 & $   -0.01393171$ & $   -0.00577056$ \\
10       & $\ps 0.00541301$ & $   -0.0653714$ & 0.00528328 & $   -0.01023858$ & $   -0.00240404$ \\
12       & $\ps 0.00425955$ & $   -0.0536435$ & 0.00474400 & $   -0.00791965$ & $   -0.00053883$ \\
14       & $\ps 0.00345867$ & $   -0.0453054$ & 0.00430652 & $   -0.00634174$ & $\ps 0.00057987$ \\
16       & $\ps 0.00287305$ & $   -0.0390806$ & 0.00394589 & $   -0.00520561$ & $\ps 0.00128963$ \\
18       & $\ps 0.00242792$ & $   -0.0342623$ & 0.00364403 & $   -0.00435276$ & $\ps 0.00175833$ \\
20       & $\ps 0.00207934$ & $   -0.0304272$ & 0.00338784 & $   -0.00369181$ & $\ps 0.00207654$ \\
22       & $\ps 0.00179983$ & $   -0.0273058$ & 0.00316769 & $   -0.00316657$ & $\ps 0.00229645$ \\
24       & $\ps 0.00157136$ & $   -0.0247185$ & 0.00297641 & $
-0.00274064$ & $\ps 0.00244972$ \\\hline $\infty$ & $   -0.00026 $
& $\ps 0.0012   $ & 0.00009    & $\ps 0.00045   $ & $\ps 0.00081 $
\\\hline
\end{tabular}
\caption{Violation of the equations of motion up to level 4.}
\label{tab:equations}
\end{table}

%%%%%%%%%%%%%%%%%%%%%%%%%%%%%%%%%%%%%%%%%%%%%%%%%%%%%%%%%%%%%%%%
\section{Summary and discussion}\label{sec:summary}
In this work, we developed a technique that allows us to compute
high level solutions in nontrivial gauges. We applied it on the
tachyon vacuum solution in the Schnabl gauge and improved the
results from \cite{Arroyo:2017iis} to level 24.

The overall behavior of energy of the Schnabl gauge solution is
similar to the Siegel gauge. In particular, it overshoots the
correct value, goes through a local minimum at level 12 and then
it turns back towards $-1$. By extrapolating the energy, we found
that the solution satisfies the Sen's first conjecture with a
precision of about $0.05\%$. This is not a bad result, but it is
worse than in the Siegel gauge. The gauge invariant $E_0$
converges towards the correct value monotonically and its infinite
level extrapolation reproduces the analytic value with a precision
of about $0.15\%$.

An intriguing question is whether the numerical solution in fact
converges to the Schnabl analytical solution. In this work, we
reach a conceivable conclusion that it does, but additional effort
will be needed to prove it with absolute certainty. At finite
levels, there is an unexpectedly big difference between the two
solutions. By employing various extrapolation techniques, we have
found that the solution moves in the right direction and we can
get much closer to the analytical solution, but the precision is
still significantly lower than in the case of the gauge invariant
observables.

In addition to straightforward comparisons of coefficients, we
tested whether the numerical solution satisfies some identities
valid for the analytical solution and the equations we projected
out during the implementation of the Schnabl gauge. Both of these
consistency checks are asymptotically satisfied quite well, which
supports the claim that the numerical solution converges to the
analytical one.

In general, extrapolations of the tachyon vacuum solution in the
Schnabl gauge are less reliable than in the Siegel gauge. Most
importantly, there is a partial instability because asymptotic
values change significantly as we add new levels. This suggests
that the Schnabl gauge solution has an unusual asymptotic
behavior. It is possible that when the solution is expanded in
$1/L$ around infinity, the coefficients have slow falloff and the
level 24 approximation is still not good enough. Another
possibility is that the numerical solution is a non-analytic
function around level infinity, and therefore one has to use a
different asymptotic expansion.

In order to understand the origin of these problems, we have made
a number of low level experiments with linear $b$-gauges. For
instance, consider a simple gauge $(b_0+\alpha b_2)\Psi=0$. In
gauges which are close to the Siegel gauge ($\alpha \rightarrow
0$), tachyon vacuum solutions behave well, while when we go
further away from the Siegel gauge, there are similar problems
with convergence as in the Schnabl gauge. The Schnabl gauge
solution is not special in this respect. Our best guess is that
the problems with convergence are caused by the fact that the
gauge condition couples states at different levels. If the
coupling is too strong, the level truncation scheme becomes less
reliable. We leave a detailed analysis of tachyon vacuum solution
in these gauges as a future research project.
\\

As another numerical experiment, we tried to find a tachyon vacuum
solution using a different approach inspired by analytical $KBc$
solutions. Consider a generic $KBc$ string field
\begin{equation}
\Psi=F_1(K)\,cF_2(K)+G_1(K)c\, G_2(K) Bc\, G_3(K).
\end{equation}
By level expansion of such string field, we find states of the
form
\begin{equation}
L_{-2k_1}^{tot}\dots L_{-2k_n}^{tot}c_{-l}|0\ra,\quad k_i>0,\
l\geq -1
\end{equation}
and
\begin{equation}
L_{-2k_1}^{tot}\dots
L_{-2k_n}^{tot}b_{-2m}c_{-l_1}c_{-l_2}|0\ra,\quad k_i>0,\
l_{1,2}\geq -1,\ m>0.
\end{equation}
We call the state space spanned by these states a {\it restricted
space}. It can be described by linear constraints similar to
(\ref{gauge-fix}) and therefore it is possible to search for SFT
solutions in the restricted space using the same techniques as for
the gauge fixing. The restricted space is not closed under the
star product, but one can hope that the projected out equations
will be satisfied in the infinite level limit.

Unfortunately, our attempts to find solutions in this setup have
not been successful. First, we tried not to impose any gauge.
However, the restricted space conditions do not fix enough
coefficients at low levels, we therefore encountered the same
difficulties as in \cite{KudrnaUniversal}. There are multiple
solutions corresponding to the tachyon vacuum and when we try to
improve them to higher levels, Newton's method fails.

Therefore, it appears necessary to impose some gauge condition in
addition to the restricted space conditions. We tested both the
Siegel and Schnabl gauge conditions, and both cases lead to
similar results. There is a unique tachyon vacuum solution at low
levels, but at higher levels, problems with the numerical
stability start to appear. Actually, Newton's method fails to
converge at level 20. Furthermore, the value of the energy
associated to the numerical solution does not appear to converge
to $-1$ and some of the projected out equations are not satisfied.

These results lead us to the conclusion that the $KBc$ algebra
cannot be consistently truncated to finite levels. The restricted
space conditions remove too many degrees of freedom at high levels
(more than the gauge fixing), therefore the remaining fields
cannot solve the SFT equation. Therefore the full analytical
expressions are needed to work with the $KBc$ algebra.

%%%%%%%%%%%%%%%%%%%%%%%%%%%%%%%%%%%%%%%%%%%%%%%%%%%%%%%%%%%%%%%%
\section*{Acknowledgements}
We would like to thank Ted Erler and Carlo Maccaferri for the
useful discussions. This research was supported partly by the
Czech Science Foundation (GA\v{C}R) grant 17-22899S and by the
European Regional Development Fund and the Czech Ministry of
Education, Youth and Sports (M\v{S}MT), project No.
CZ.02.1.01/0.0/0.0/15\_003/0000437. Computational resources were
provided by the CESNET LM2015042 and the CERIT Scientific Cloud
LM2015085, provided under the programme "Projects of Large
Research, Development, and Innovations Infrastructures.

\newpage
\appendix
\setcounter{equation}{0}
\def\thesection{\Alph{section}}
\renewcommand{\theequation}{\Alph{section}.\arabic{equation}}

\section{Extrapolations to infinite level} \label{sec:extrapolations}
In this appendix, we provide general rules for extrapolations
applied to the data of various quantities.

Suppose we have a set of data points for some quantity from level
$L_{min}$ to $L_{max}$ \footnote{It is assumed that we have data
points at even levels only. If we have data at odd levels (or
possibly at fractional levels), the data points must be divided
into groups, where the members of each group differ by two levels,
and each group must be extrapolated separately, see
\cite{KudrnaThesis} for more details. In the case of Ellwood
invariants, the interval is four levels, so there are always at
least two groups.}. In order to perform extrapolations, we will be
required to fit this set of data points using functions of the
form \cite{KudrnaThesis}\cite{Kudrna:2016ack}
\begin{align}
\label{1overLI} q^{(L_{min},L_{max})}_M(L)=a_0 + \sum_{n=1}^{M}
\frac{a_n}{L^n},
\end{align}
where the order of the fit $M$ can vary from $M_{min}=1$ to its
maximal value $M_{max}=(L_{max}-L_{min})/2$. We denote this
maximal value as $M_{max}\equiv N$. Note that the value of $N$ is
equal to the number of used data points minus 1.  The coefficients
$a_0,a_1,\cdots,a_M$ are determined by minimizing the sum of the
squares of the residuals of the data points from the function
(\ref{1overLI}). Namely, we use the least squares fitting
technique. In practice, we use the function NonlinearModelFit in
Mathematica.

In the case of the Siegel gauge, we usually use all available data
points and maximal order fits to perform extrapolations. This
procedure works very well, see \cite{Gaiotto:2002wy} and also
\cite{KudrnaThesis}. Let us analyze whether this approach also
works in the Schnabl gauge. A reliable extrapolation procedure
should be applicable to multiple quantities and it should satisfy
the following properties:
\begin{enumerate}
  \item It should approximate the data well.
  \item It should be stable, namely its result should not change much if we add/remove
   a data point or change the order of the fit.
  \item It should reproduce the analytical expected values.
\end{enumerate}

To test the first point, for a set of data corresponding to some
quantity, we can check whether we can predict level 24 data using
lower level data. We consider all possible data sets from level
$L_{min}$ to $L_{max}$, which go from 2 to 22, and fits with all
possible orders $M$.

To determine which of the fits (\ref{1overLI}) provides the best
prediction for the value of the quantity at level $L=24$, we
evaluate $| \Delta | \equiv \left|
q^{(L_{min},L_{max})}_M(24)-q(24)\right|$, which represents the
absolute value of the residuals of the predicted value of the
quantity from its actual value $q(24)$ that has been obtained by
direct level truncation computations.

In table \ref{tab:experiment2}, we show what are the best three
predictions of level 24 data for several different quantities,
which include the energy, some string field coefficients and some
out-of-gauge equations. We can see that all the best fits use the
maximal possible order and all or almost all data points.
Therefore the procedure used in the Siegel gauge should work in
the Schnabl gauge too. Note that these fits are interpolation
functions that pass through the data points. This result suggests
that the level truncation data lie on some smooth curve and they
do not contain random fluctuations like experimental data (for
instance, due to noise effects).

\begin{table}[!]
\centering
\begin{tabular}{|cccc|cccc|}\hline
$L_{min}$ & $L_{max}$ & $M$ & $|\Delta|$ & $L_{min}$ & $L_{max}$ &
$M$ & $|\Delta|$ \\\hline \multicolumn{4}{|c|}{Energy}         &
\multicolumn{4}{c|}{$t$}           \\\hline
 4 & 22 & 9  & $4.27\times 10^{-10}$ & 2 & 22 & 10 & $2.15\times 10^{-8}$  \\
 6 & 22 & 8  & $8.38\times 10^{-10}$ & 4 & 22 & 9  & $2.38\times 10^{-8}$  \\
 2 & 22 & 10 & $9.17\times 10^{-10}$ & 6 & 22 & 8  & $2.94\times 10^{-8}$  \\\hline\hline
\multicolumn{4}{|c|}{$v$}            &   \multicolumn{4}{c|}{$u$}
\\\hline
 2 & 22 & 10  & $3.47\times 10^{-8}$ & 2 & 22 & 10 & $9.55\times 10^{-8}$ \\
 4 & 22 & 9   & $3.92\times 10^{-8}$ & 4 & 22 & 9  & $1.08\times 10^{-7}$ \\
 6 & 22 & 8   & $5.04\times 10^{-8}$ & 6 & 22 & 8  & $1.39\times 10^{-7}$ \\\hline\hline
\multicolumn{4}{|c|}{$f_w$}          &
\multicolumn{4}{c|}{$f_{w_1}$}    \\\hline
 2 & 22 & 10 & $3.79\times 10^{-9}$  & 2 & 22 & 10 & $1.68\times 10^{-8}$ \\
 4 & 22 & 9  & $4.20\times 10^{-9}$  & 4 & 22 & 9  & $1.79\times 10^{-8}$ \\
 6 & 22 & 8  & $5.21\times 10^{-9}$  & 6 & 22 & 8  & $2.08\times 10^{-8}$ \\\hline
\end{tabular}
\caption{Predictions for level 24 data for various quantities in
the Schnabl gauge. We consider all possible data sets from level
$L_{min}$ to $L_{max}$, which go from 2 to 22, and fits with all
possible orders $M$. We show the best three results for all
quantities, which suggest that it is best to use maximal order
fits and as much data points as possible.} \label{tab:experiment2}
\end{table}

Regarding the stability of extrapolations, we observe that the
results are worse than in the Siegel gauge and there is only a
partial stability. For instance, see table \ref{tab:EnergyN},
where we show asymptotic values of the energy that were computed
using extrapolations of the form $E_N^{(2,2N+2)}(L)$. Although the
results do not change much when we change $N$ by one unit, the
differences are bigger than in Siegel gauge. A similar behavior
can be seen in the case of string field coefficients in table
\ref{tab:coefN1}.

We have not found any fits for the energy, as well as for the
string field coefficients, that are free of this stability issue.
This probably suggests that the numerical solution in the Schnabl
gauge has some unusual asymptotic behavior.

\begin{table}[!]
\centering
\begin{tabular}{|cccc|cccc|}\hline
$L_{min}$ & $L_{max}$ & $M$ & $\infty$ & $L_{min}$ & $L_{max}$ &
$M$ & $\infty$ \\\hline \multicolumn{4}{|c|}{$f_{w}$}         &
\multicolumn{4}{c|}{$f_{w_1}$}           \\\hline
 4 & 24 & 10  & $-0.00026$ &  4 & 8  & 2  &  0.00091 \\
 6 & 24 & 9   & $-0.00027$ &  2 & 12 & 5  &  0.00098 \\
 8 & 24 & 8   & $-0.00029$ &  4 & 24 & 10 &  0.00119 \\\hline\hline
\multicolumn{4}{|c|}{$f_{w_2}$}         &
\multicolumn{4}{c|}{$f_{w_3}$}           \\\hline
 4 & 24 & 10  & 0.000091 & 2 & 14 & 5  & $\ps 0.00002$ \\
 6 & 24 & 9   & 0.000094 & 2 & 16 & 5  & $\ps  0.00031$ \\
 8 & 24 & 8   & 0.000101 & 2 & 12 & 5  & $ -0.00044 $ \\\hline
\end{tabular}
\caption{The best three extrapolations for some out-of-gauge
equations in the Schnabl gauge. We considered all possible data
sets from level $L_{min}$ to $L_{max}$, which go from 2 to 24, and
fits of all possible orders $M$. Overall, the best choice for
these extrapolations of these equations is to use the data from
levels 4 to 24 and the maximal order $N=10$. There are sometimes
better agreements (for example for $f_{w_3}$, where the suggested
fit is only the 4th best), but these are clearly coincidental.}
\label{resultsoutgauge1}
\end{table}

In the case of other quantities, like the gauge invariant overlap
and the out-of-gauge equations, we observed that their
extrapolated values are more stable than those observed in the
case of the energy. For the out-of-gauge equations, the best
results are generally obtained by maximal order fits using data
from level 4, see table \ref{resultsoutgauge1}.

Finally, regarding the third point, it turns out that the energy
and, even more significantly, string field coefficients have
lesser precision than we would like and there are essential
differences between the extrapolated and the analytical expected
values. Nevertheless, we observe that the extrapolations clearly
move in some direction as we increase the order $N$. Therefore we
can try to make a second, $N \rar \inf$, extrapolation which
potentially allows us to improve the results. So let us establish
the procedure to determine this $N\rar \inf$ extrapolation.

Suppose we have data points from level $L_{min}$ to $2N+L_{min}$.
As a first step, we construct functions of the form
$q^{(L_{min},2N+L_{min})}_{N}(L)$ and evaluate their asymptotic
values, namely we compute $q_N\equiv \lim_{L \rar \infty}
q^{(L_{min},2N+L_{min})}_{N}(L)$.

As a second step, by means of polynomials in $1/N$ of order $M$
\begin{align}\label{1overNI1}
r_0 + \sum_{n=1}^{M} \frac{r_n}{N^n},
\end{align}
we fit the set of data points $\{q_1,q_2,\cdots,q_N\}$, where the
coefficients $r_0,r_1,\cdots,r_M$ are again determined using the
least squares fitting technique. Finally, we compute the
asymptotic value of (\ref{1overNI1}) when $N\rar \inf$. In this
extrapolation procedure, it turns out that it is necessary to use
less than maximal order fits, because fits with high $M$ are
unstable.

This second $N\rar \inf$ extrapolation procedure works quite well
for the string field coefficients from section
\ref{sec:coefficients}, for which we find that the best results
are usually given by fits with $M=5$. While in the case of the
energy and out-of-gauge equations, the differences between fits of
different order $M$ are of similar size as the precision of
original $L \rar \infty$ extrapolations.

\section{Search for other solutions}\label{sec:other sol}
Following \cite{KudrnaUniversal}, we also attempted to search for
other Schnabl gauge solutions different from the tachyon vacuum.
We used the homotopy continuation method adapted to Schnabl gauge
to find all solutions of the equations of motion at low levels,
which serve as seeds for Newton's method. We managed to get to
level 6 with the twist even condition imposed and to level 5
without it. As in the Siegel gauge, we found several millions of
solutions, however, most of them have $|E|\gg 1$, which means that
there is only a small probability that they would represent some
physical configuration. Subsequently, we took several solutions
close to the perturbative vacuum and improved them to higher
levels. As in the Siegel gauge, we have found several solutions
which are more or less stable in the level truncation scheme,
nonetheless only two of them behave sufficiently well to motivate
a closer attention. Both solutions are twist even and they appear
at level 4.

The properties of the first solution are summarized in table
\ref{tab:other sol 1}. The solution appears to be an analogue of
the "double brane" found in \cite{KudrnaUniversal}, but
 it behaves asymptotically similarly to the "half brane" solution. The
extrapolated values of its energy and $E_0$ are non-integers,
exhibiting large imaginary parts, so this solution is most likely
not physical.

The second solution, which is shown in table \ref{tab:other sol
2}, behaves slightly better. It is real and the extrapolated
values of its energy and the gauge invariant are close to 0
\footnote{The energy and $E_0$ in this appendix are normalized
according to the convention adopted in reference
\cite{KudrnaUniversal}. Recall that in the conventions of this
work, the analytical value of the vacuum energy and the gauge
invariant $E_0$ are equal to $-1$ and $0$ respectively.}.
Therefore, it is possible that this solution is gauge equivalent
to the tachyon vacuum. However, the precision of its energy is
quite low and the first out-of-gauge equation is not exactly
satisfied, so it is possible that this solution is a relict of the
level truncation approach as well.

\begin{table}
\centering
\begin{tabular}{|c|lll|}\hline
$L$      & $\ps$Energy              & $E_0$                & $\ps
f_w$                 \\\hline
4        & $\ps 3.282522-1.78319 i$ & $1.161906-1.69528 i$ & $   -0.306633-1.425542 i$ \\
6        & $\ps 1.568079-1.64296 i$ & $0.742676-1.23355 i$ & $   -0.397242-0.606541 i$ \\
8        & $\ps 1.098815-1.33103 i$ & $0.596923-1.08489 i$ & $   -0.329192-0.374098 i$ \\
10       & $   -0.912291-1.12913 i$ & $0.549851-0.95310 i$ & $   -0.279083-0.274678 i$ \\
12       & $   -0.819925-0.99526 i$ & $0.512582-0.88831 i$ & $   -0.244791-0.221596 i$ \\
14       & $   -0.767364-0.90108 i$ & $0.499967-0.82554 i$ & $   -0.220437-0.189119 i$ \\
16       & $   -0.734482-0.83139 i$ & $0.484460-0.78922 i$ & $   -0.202383-0.167350 i$ \\
18       & $   -0.712456-0.77768 i$ & $0.480014-0.75187 i$ & $   -0.188498-0.151777 i$ \\
20       & $   -0.696923-0.73497 i$ & $0.471866-0.72834 i$ & $   -0.177490-0.140085 i$ \\
22       & $   -0.685523-0.70012 i$ & $0.470194-0.70323 i$ & $   -0.168545-0.130974 i$ \\
24       & $   -0.676885-0.67110 i$ & $0.465287-0.68656 i$ & $
-0.161124-0.123662 i$ \\\hline $\infty$ & $   -0.630   -0.318   i$
& $0.456   -0.448   i$ & $   -0.073   -0.048    i$ \\\hline
\end{tabular}
\caption{Energy, Ellwood invariant and the first out-of-gauge
equation for a "half brane" solution.} \label{tab:other sol 1}
\end{table}

\begin{table}[!]
\centering
\begin{tabular}{|c|lll|}\hline
$L$      & $\ps$Energy   & $\ps E_0$     & $\ps f_w$      \\\hline
4        & $   -25.8203$ & $   -1.526687$ & $\ps 0.110970$ \\
6        & $   -11.0589$ & $   -1.027420$ & $\ps 0.126293$ \\
8        & $   -6.77329$ & $   -0.757402$ & $   -0.021528$ \\
10       & $   -4.89103$ & $   -0.617817$ & $   -0.090068$ \\
12       & $   -3.85610$ & $   -0.524986$ & $   -0.123871$ \\
14       & $   -3.20556$ & $   -0.461964$ & $   -0.142606$ \\
16       & $   -2.75936$ & $   -0.414177$ & $   -0.153968$ \\
18       & $   -2.43407$ & $   -0.377513$ & $   -0.161320$ \\
20       & $   -2.18609$ & $   -0.347829$ & $   -0.166303$ \\
22       & $   -1.99047$ & $   -0.323443$ & $   -0.169790$ \\
24       & $   -1.83196$ & $   -0.302941$ & $   -0.172285$
\\\hline $\infty$ & $   -0.19   $ & $   -0.054   $ & $   -0.15
$ \\\hline
\end{tabular}
\caption{Energy, Ellwood invariant and the first out-of-gauge
equation for a possible gauge copy of the tachyon vacuum
solution.} \label{tab:other sol 2}
\end{table}

\newpage
%%%%%%%%%%%%%%%%%%%%%%%%%%%%%%%%%%%%%%%%%%%%%%%%%%%%%%%%%%%%%%%%

\end{document}